\documentclass{article}
\usepackage[utf8]{inputenc}
\usepackage[english]{babel}
\usepackage{graphicx}
\usepackage{slashed}
\usepackage{hyperref} 
\usepackage{geometry}
\usepackage{amsmath}
\usepackage{amssymb}
\usepackage{amsfonts}
\usepackage{multicol}
\usepackage{float}
\usepackage{caption}
\usepackage{subcaption}
\usepackage{cite}
\usepackage{amsthm}
\usepackage[utf8]{inputenc}
\usepackage{xcolor}
\title{\bf Theoretical bounds on dark matter masses}
\author{Xavier Calmet\thanks{E-mail: X.Calmet@sussex.ac.uk}$\ $ 
and Folkert~Kuipers\thanks{E-mail: F.Kuipers@sussex.ac.uk} 
\\
\\
{\em Department of Physics and Astronomy, University of Sussex,}\\ 
{\em Brighton, BN1 9QH, United Kingdom}
}

\begin{document}
\maketitle
%\vspace{2cm}
%
\begin{abstract}
In this letter, we show that quantum gravity leads to lower and upper bounds on the masses of dark matter candidates. These bounds depend on the spins of the dark matter candidates and the nature of interactions in the dark matter sector. For example, for singlet scalar dark matter, we find a mass range $10^{-3}{\rm eV} \lesssim m_\phi \lesssim 10^{7}{\rm eV}$. The lower bound comes from limits on fifth force type interactions and the upper bound from the lifetime of the dark matter candidate.
\end{abstract}
\flushbottom
%\raggedbottom
%
\thispagestyle{empty}
\pagebreak
\pagenumbering{arabic}

There is overwhelming evidence that most of the matter in our universe is dark and cannot be described by the Standard Model of particle physics. The case for the existence of dark matter is strong because it comes from astrophysical and cosmological observations made on different scales and times in our universe. For example, the cosmic microwave background or galaxy rotation curves involve very different physics and eras in the evolution of our universe but they both require that about 75\% of the matter content of the universe consists of cold, non-baryonic, dark matter.

From a theoretical point of view, very little is known of the nature of dark matter.  We know that there is no viable candidate in the Standard Model of particle physics. There are basically three different approaches. The first approach consists in introducing a new particle stable enough over the lifetime of the universe which couples at most extremely weakly to the photon so that it remains dark enough. A typical example of such a particle would be a weakly interacting massive particle (WIMP), see e.g. \cite{Roszkowski:2017nbc} for a review. The second one consists in modifying gravity see e.g. \cite{Milgrom:1983ca,Bekenstein:2004ne,Moffat:2005si,Capozziello:2011et}, but it is difficult to construct a proper model and even when that is case, it has been argued \cite{Calmet:2017voc} that this approach is identical to the first one with the caveat that the new field is only coupled gravitationally to the Standard Model particles. Finally, one could hope that some massive astrophysical compact halo objects (MACHOs) such as primordial black holes \cite{Green:2020jor} could explain the missing matter without having to modify the Standard Model or General Relativity. Alas, this solution to the dark matter problem, while beautifully simple and minimalistic as it does not require new physics beyond the Standard Model or General Relativity, does not appear to be relevant to Nature, see e.g. \cite{Freese:1999ge}.

If we accept that new physics is required to address the missing matter problem, we are faced with a huge theoretical challenge as we have very little information about the nature of the dark matter particle or particles. We do not know their spins, masses, self-interactions or couplings to the Standard Model particles. Galaxy formation simulations seem to prefer cold, i.e. non-relativistic, dark matter. The interactions of dark matter particles with that of Standard Model or dark matter self-interactions must be weak see e.g. \cite{Tulin:2017ara} for a review.

Fortunately, quantum gravity can provide some guidance on the allowed parameter range for a given dark matter candidate. The reason for this is simple. In general, quantum gravitational effects will lead to a decay of any dark matter candidate that is not protected by Lorentz invariance or a gauge symmetry from decaying. Furthermore, gravity is universal, it will thus couple to all forms of matter and it will create portals between the Standard Model and any hidden sector. While these decays will be suppressed by powers of the Planck mass, they will still lead to an upper bound on dark matter particles given the large age of our universe. Furthermore, if the dark matter particles are light, the same quantum gravitational effects will lead to fifth force type interactions and these interactions are bounded by limits coming from the E\" ot-Wash experiment \cite{Kapner:2006si,Hoyle:2004cw,Adelberger:2006dh,Leefer:2016xfu,Braginsky1972,Smith:1999cr,Schlamminger:2007ht,Adelberger:2009zz,Zhou:2015pna}. Finally, there is a well known lower bound coming from quantum mechanics and more specifically the spin-statistics theorem which applies to fermionic dark matter candidate. This last bound depends on the dark matter profile. Putting all these bounds together, we obtain tight mass ranges for scalar, pseudo-scalar, spin 1/2 and spin 2 dark matter particles which are gauge singlets. These bounds can be relaxed if the fields describing these particles are gauged, we however note that there are fairly tight constraints on the strength of the interactions in the dark matter sector. Finally, we argue that spin-1 vector dark matter particles are less constrained by quantum gravity, because of the chiral nature of the fermions in the Standard Model.

We consider local operators that are generated by non-perturbative quantum gravity effects (see e.g. \cite{Calmet:2019jyz,Calmet:2019frv,Calmet:2020rpx,Holman:1992us,Barr:1992qq,Calmet:2014lga,Giddings:1988cx,Coleman:1988tj,Gilbert:1989nq}):
\begin{equation}
	O_1 = \frac{c_\phi}{M_{\rm P}}  \, \phi \, F_{\mu\nu}F^{\mu\nu},
\end{equation}
where $M_{\rm P}=2.4 \times 10^{18}$ GeV is the reduced Planck scale, $\phi$ is the scalar dark matter field, and $F_{\mu\nu}$ is the electromagnetic field tensor.  We note that there are solid arguments showing that the Wilson coefficient $c_1$ is of order one \cite{Calmet:2020rpx}.

The results from the E\"{o}t-Wash torsion pendulum experiment that searches for fifth forces \cite{Kapner:2006si,Hoyle:2004cw,Adelberger:2006dh,Leefer:2016xfu,Braginsky1972,Smith:1999cr,Schlamminger:2007ht,Adelberger:2009zz,Zhou:2015pna}  imply that $m_\phi \gtrsim 10^{-3}\, {\rm eV}$ \cite{Calmet:2019jyz,Calmet:2019frv,Calmet:2020rpx}.  The same operator can lead to the decay of the dark matter scalar \cite{Calmet:2009uz,Mambrini:2015sia} with a decay width $\Gamma \sim  m_\phi^3/(4 \pi M_P^2)$ and lead to an upper bound $m_\phi \lesssim 10^{7}{\rm eV}$ from the requirement that the dark matter candidate lives long enough to still be present in today's universe.  Quantum gravity thus enables to restrict the mass of any singlet scalar particle to be in the range:
\begin{equation}
10^{-3}{\rm eV} \lesssim m_\phi \lesssim 10^{7}{\rm eV},
\end{equation}
independently of its potential non-gravitational couplings to Standard Model particles or self-interactions. Note that these bounds would not apply to a gauged scalar field as only dimension six operators would be generated by quantum gravity. In that case, one has $m_\phi \gtrsim 10^{-22}{\rm eV}$  \cite{Calmet:2020rpx}, and the upper bound disappears \footnote{Note that some readers may be worried about the naturalness of very light scalars. We take an agnostic approach and simply derive bounds from quantum gravity assuming that such light scalars exist.}. 

The same bound applies to  the mass of a pseudo-scalar dark matter candidate, an axion like particle, $a$ if quantum gravity violates parity (and time reversal invariance) \cite{Calmet:2020rpx}
\begin{equation}
10^{-3}{\rm eV} \lesssim m_a \lesssim 10^{7}{\rm eV}.
\end{equation}

On the other hand, if quantum gravity preserves parity,  we have to consider the operator
\begin{equation}\label{LinCoup}
	O_a= \frac{c_a}{M_{\rm P}}  \, a\, \tilde F_{\mu\nu}F^{\mu\nu}.
\end{equation}
For an axion-like-particle, we then find \cite{Calmet:2020rpx,Mambrini:2015sia}
\begin{equation}
 10^{-21}{\rm eV} \lesssim m_a \lesssim 10^{7}{\rm eV},
\end{equation}
for parity conserving quantum gravity. The upper bound comes from the requirement that the particle is long-lived in comparison to the age of the universe and the lower bound is derived from magnetometry searches \cite{Calmet:2020rpx,Abel:2017rtm}.

For spin $1/2$ fermions $\psi$, quantum gravity leads to an upper bound on the mass of the dark matter candidate \cite{Calmet:2009uz,Mambrini:2015sia,Boucenna:2012rc}  as it could decay to the Standard Model fields, while a lower bound comes from the Pauli exclusion principle. We consider the operator \cite{Calmet:2009uz,Mambrini:2015sia}:
\begin{equation}
	O_\psi = \frac{c_\psi}{M_{\rm P}}  \, \bar \psi \tilde H^\dagger \slashed{D} L,
\end{equation}
where $H$ is the Higgs doublet of the Standard Model with $\tilde H= -i \sigma_2 H^\ast$. This operator implies that the singlet right-handed fermion $\psi$ can decay to an off-shell $Z$ boson and a neutrino, the $Z$ boson then decays to two light fermions. Requiring that the fermion singlet lives long enough to still be present today imposes an upper bound on its mass. One finds $m_\psi < 10^{10}{\rm eV}$  using $\Gamma=v^2 G_F^2 m_\psi^5/(192 \pi^3 M_P^2)$ where $G_F$ is the Fermi constant and $v=246$ GeV the electroweak vacuum expectation value.

Since fermions cannot be in the same state, only a limited amount of fermions can be present in a galaxy with momenta below the escape velocity. Together with the assumption that the fermions must account for the observed dark matter density in a typical galaxy this leads to a lower bound on the mass of the fermions\cite{Tremaine:1979we,DiPaolo:2017geq,Savchenko:2019qnn}. The bounds on the mass of the dark fermion are then given by 
\begin{equation}
	10^{2}{\rm eV} \lesssim m_\psi \lesssim 10^{10}{\rm eV}.
\end{equation}
The lower bound holds for the Standard Model, but it can be relaxed by assuming multicomponent dark matter \cite{Davoudiasl:2020uig}.

We now consider a vector boson dark matter $V^\mu$. The well studied dimension four operator $F^{\mu\nu}B_{\mu\nu}$, where $F^{\mu\nu}$ is the field strength of the hypercharge photon of the Standard Model and $B_{\mu\nu}$ that of the dark photon, while generated by quantum gravity, is expected to be exponentially suppressed  \cite{Calmet:2009uz,Calmet:2020rpx}. Within the Standard Model, the only dimension five gauge invariant operator is given by  $c_{V,5} M^{-1}_{\rm P}  \, V^\mu (\bar \psi_R i \tilde H^\dagger \gamma_\mu L )$
but after electroweak symmetry breaking, this simply accounts for a shift of the photon field. The next operators are of mass dimension 6
$c_{V,6} M^{-2}_{\rm P}  V_\mu (H^\dagger D_\nu H) F^{\mu\nu}$ or $M^{-2}_{\rm P} ( \bar \psi \sigma_{\mu\nu} \tilde H^\dagger \slashed{D} L) B^{\mu\nu}$.  These operators lead to dimension five operators after electroweak symmetry breaking but there is a chiral suppression $v/M_P$. The only useful dimension five operator involves the production of a graviton $h_{\mu\nu}$
\begin{equation}
	O_V = \frac{c_V}{M_{\rm P}}  h_{\mu\alpha} F^{\mu}_{\ \nu} B^{\nu \alpha}\, ,
\end{equation}
which enables the decay of a vector dark matter to a photon and a graviton. This operator exists in the Standard Model with the vector boson replaced by a $Z$-boson \cite{Nieves:2005ti}. It is straightforward to estimate the decay width of the $V$ boson, one finds $\Gamma \sim c_V^2 m_V^3/M_P^2$ and we can thus find an upper bound on the mass of a vector dark matter particle from the requirement that it is still around in today's universe. We find $m_V<10^7$ eV. We can get a lower bound on its mass if we assume that all of dark matter is described by a vector particle. 
As for a scalar field, see e.g. \cite{Stadnik:2018sas} for a recent review, the requirement that the boson's de Broglie wavelength does not exceed the dark matter halo size of the smallest dwarf galaxies gives a lower bound on its mass $m_V > 10^{-22}$ eV. We thus find
\begin{equation}
	10^{-22}{\rm eV} \lesssim m_V \lesssim 10^{7}{\rm eV}.
\end{equation}

Using the results developed in \cite{Calmet:2019frv}, it is straightforward to see that for a massive spin-2 field dark matter field, one obtains similar bounds for its mass to that of a singlet scalar field dark matter candidate:
\begin{equation}
10^{-3}{\rm eV} \lesssim m_2 \lesssim 10^{7}{\rm eV}.
\end{equation}

In this letter, we have shown that a few very well motivated theoretical concepts based on quantum gravity and the spin-statistics theorem enable to constrain the masses of low spin dark matter candidates. Quantum gravity generates operators that will lead to a decay of all dark matter candidates that are represented by fields that are not gauged or prevented by Lorentz invariance from decaying to Standard Model particles. This lead to an upper bound on their masses. If these dark matter candidates are bosons, they will mediate a fifth force and we can apply bounds from the E\"ot-Wash experiment which provide a lower bound on their masses. In the case of fermion dark matter candidates, the lower bound comes from the spin-statistics theorem. 

Our bounds are derived assuming the worst case scenario for quantum gravity, namely that it has only one scale and that this scale is the traditional reduced Planck scale i.e. $2.4 \times 10^{18}$ GeV. In other words, we assumed that quantum gravity is as weak as possible. Our bounds become much more stringent if the effective scale of quantum gravity is below $2.4 \times 10^{18}$ GeV as it is the case in models with large extra-dimensions where it could be in the TeV region or if there is another infra-red cutoff that is below the reduced Planck mass as it is the case in some specific models of quantum gravity see, e.g., \cite{Eichhorn:2020mte,Minic:2020oho,Strominger:2017zoo}.

We would like to stress that our bounds are orders of magnitude estimates. We argue that because we are dealing with non-perturbative quantum gravity, the only relevant coupling constant should be the Planck mass. It is however conceivable that there is a further suppression of some of the Wilson coefficients which could involve coupling constants of the Standard Model. For example, $c_\phi$ could contain a factor $g^2/(4\pi)$ where $g$ is the hyperfine coupling constant of the $U(1)$ group of the Standard Model or $c_\psi$ could be proportional to the electron Yukawa coupling which is of the order of $10^{-5}$. Clearly, this would impact our bounds. Here, we made the strong assumption that the dimension five operators are of pure quantum gravitational origin. 

Finally, as explained already, we emphasize that these bounds will not apply to hidden sector fields that are gauged under some gauge symmetry whether this is a continuous or discrete gauge symmetry \cite{Krauss:1988zc,Banks:2010zn}. For gauged fields, dimension 5 operators will not be generated directly, one would expect dimension 6 or higher operators. For a gauged scalar field $\Phi$  for example, one has $M_{\rm P}^{-2}\Phi \cdot \Phi \, F_{\mu\nu}F^{\mu\nu}$ in which case we can only exclude masses $m_\Phi\lesssim 10^{-22}\, {\rm eV}$.  Dimension 5 operators, if they exist, would have a further suppression if they are generated from a higher dimensional operators. For example, if $\Phi$ has some none-vanishing expectation value $v_\Phi$ in the TeV region, the resulting dimension five operator $v_\Phi M_{\rm P}^{-2}\phi\, F_{\mu\nu}F^{\mu\nu}$ would be suppressed by a factor $v_\Phi/M_{\rm P}\sim 10^{-16}$. A similar suppression would be generated in models with a discrete gauge symmetry. Such a suppression would open up the allowed mass range for dark matter candidates. The situation is similar for a complex scalar dark matter, see e.g. \cite{Boehm:2020wbt}, which carries a charge: quantum gravity would form operators of the type $M_{\rm P}^{-2}\ \phi^\star \phi \, O_{SM}$ (where $O_{SM}$ are operators build with fields of the Standard Model) which would be at least of dimension 6, if the complex scalar field is gauged. If it is a discrete or global symmetry, one would expect that quantum gravity breaks this symmetry. One would then obtain operators of the type $M_{\rm P}^{-1}\ \phi \, O_{SM}$ and our bounds would apply. This is particularly important in the case of WIMPs, which are largely excluded by our bounds, if the WIMP is a gauge singlet. Our bounds can be avoided if one gauges WIMPs. Clearly the origin of the dimension five operators that we have discussed in this letter is model dependent and one needs to verify on the case-by-case whether such operators will be generated in a specific dark matter model.

\section*{Acknowledgments}
We would like to thank Stephen Barr and Mark Wise for useful suggestions. The work of X.C.~is supported in part  by the Science and Technology Facilities Council (grants numbers ST/T00102X/1, ST/T006048/1 and ST/S002227/1). The work of F.K.~is supported by a doctoral studentship of the Science and Technology Facilities Council.

\end{document}